\documentclass[12pt]{article}

\usepackage{amsfonts}

\begin{document}

\input amssym.def

\input amssym

\centerline{\Large \bf  Electronic Fock Space as Associative}

\centerline{\Large \bf Superalgebra}

\centerline {A. I. Panin}

\centerline{ \sl Chemistry Department, St.-Petersburg State University,}

\centerline {University prospect 26, St.-Petersburg 198504, Russia }

\centerline { e-mail: andrej@AP2707.spb.edu }

\bigbreak

{\bf ABSTRACT: }{\small New algebraic structure on electronic Fock
space  is studied in detail. This structure  is defined in terms
of a certain multiplication of many electron wave functions and
has close interrelation  with coupled cluster  and similar
approaches. Its study  clarifies and simplifies the mathematical
backgrounds of these approaches. And even more, it leads to many
relations that would be very difficult to derive using
conventional technique. Formulas for action of the
creation-annihilation operators on products of state vectors are
derived. Explicit expressions for action of simplest
particle-conserving products of the creation-annihilation
operators on powers of state vectors are given. General scheme of
parametrization of representable density operators of arbitrary
order is presented.}

\bigbreak {\bf Key words: }{\small Fock space, commutative and
skew-commutative algebras, configuration interaction, coupled
cluster approach, density operators}

\bigbreak

\hrule
\bigbreak

{\Large \bf Introduction}

\bigbreak

In our previous paper \cite {Panin-1} new structure of associative
and skew-commuta-tive algebra (which is called superalgebra
structure by physicists) on electronic Fock space was defined.
This structure presupposes initial selection of some Hartree-Fock
(HF) vacuum state and is actually introduced on the so-called
pointed Fock space. Selection of HF vacuum states corresponding to
different numbers of electrons results in different superalgebra
structures. In particular, selection of the 'absolute' vacuum
state (with no particles) leads to the standard Grassmann algebra
with the classic wedge product. If some $p-$electron determinant
is selected as the HF vacuum state then $p-$electron sector of the
Fock space is closed with respect to the multiplication introduced
and it turns out that this sector becomes a commutative and
associative algebra. The structure of commutative algebra thus
defined proves to be very closely related to the coupled cluster
(CC) \cite {Bartlett-1}-\cite {Crawford} and quadratic
configuration interaction (QCI) approaches \cite {Pople}.

In present paper we continue our study of new algebraic structure introduced on
the electronic Fock space.

In Section I necessary basic definitions are given.

In Section II the action of the  creation-annihilation operators on arbitrary products
of state vectors (not necessarily corresponding the the same number of
electrons) is studied.

Section III is dedicated to investigation of commutative algebra structure on
the $p-$electron sector of the Fock space. In particular, general formulas
obtained in Section II are applied to arbitrary powers of $p-$electron
state vectors. General polynomial parametrization of $p-$electron states,
embracing both CC and QCI parametrizations, is described. Compact expression
for the first order CC density matrix is derived.

In Section IV non-standard realizations of electronic Fock space
suggested in \cite {Panin-2,Panin-3} and based on the notion of
$(p,q)-$vector are discussed. It is demonstrated that these
realizations are easily defined with the aid of the so-called
Hodge isomorphism \cite {Hodge}. Since the notion of
$(p,q)-$vector is rather unusual for quantum chemists, formal
definitions are accompanied by a number of simple examples. Close
relation of $(p,q)-$vectors and reduced density operators of order
$q$ is discussed. It is shown that non-standard realizations of
the Fock space give rare possibility to construct parametric
representable density operators in uniform and general way.

\bigbreak \bigbreak

{\Large \bf Basic Definitions} \bigbreak \bigbreak

\bigbreak \bigbreak

Let $N=\{1,2,\ldots,n\}$ be the molecular spin-orbital (MSO) index
set. For arbitrary $K=k_1<k_2<\ldots<k_s\subset N$ let us put
$${\Delta}_K=\{1,2,\ldots,k_1\}\Delta\ldots \Delta \{1,2,\ldots,
k_s\}\eqno(1)$$ where $\Delta $ is associative set-theoretical
operation (symmetric difference) defined on the set ${\cal P}(N)$
of all subsets of $N$ as
$$R\Delta S=(R\cup S)\backslash (R\cap S).$$
For example, if $K=\{2,4,7\}$ then
$${\Delta}_K=\{1,2\}\Delta \{1,2,3,4\}\Delta \{1,2,3,4,5,6,7\}=
\{1,2,5,6,7\}.$$ Among numerous relations involving operation
${\Delta}_K$ the equality
$$|K\cap {\Delta}_ R|+|R\cap {\Delta}_ K|\equiv |K||R|+|K\cap R|{\ } (mod{\ } 2) \eqno(2)$$
that is an immediate consequence of definition (1), is of primary
importance for further analysis.

 Finite-dimensional Fock space
${\cal F}_N$ generated by some orthonormal set $\{{\psi}_i\}_{i\in
N}$ of MSOs is constructed as follows. The basic space is the
one-electron sector of the Fock space
$${\cal F}_{N,1}=\bigoplus_{i\in N}{\mathbb C}{\psi}_i\eqno(3)$$
The direct sum of exterior powers of ${\cal F}_{N,1}$ is, by
definition, the electronic Fock space
$${\cal F}_N=\bigoplus_{p=0}^{n}\bigwedge^p {\cal F}_{N,1}\eqno(4)$$
where
$${\cal F}_{N,0}=\bigwedge^0 {\cal F}_{N,1}={\mathbb C}|\emptyset \rangle , \eqno(5)$$
${\mathbb C}$ is the field of complex numbers, $|\emptyset \rangle
$ is the so-called 'absolute' vacuum vector being nothing more
than a special notation for the unit of ${\mathbb C}$. Sector of
the Fock space corresponding to $p-$electron system is just the
$p-$th exterior power in the sum (4). It is spanned by ${n\choose
p}$ vectors
$$|R\rangle ={\psi}_{r_1}\wedge \ldots \wedge {\psi}_{r_p}\eqno(6)$$
that are called 'determinants' by quantum chemists, and by
factorable $p-$vec-tors by mathematicians. In Eq.(6) we assume
that $R=r_1<\ldots<r_p \subset N$.

{\it Basis determinants (factorable multivectors) spanning the
Fock space will be labelled by subsets of the index set $N$ and
all sign conventions connected with their representation as the
wedge products of ordered spin-orbitals will be included in the
definition of the creation-annihilation operators. This
representation is very close to the so-called space of occupation
numbers where each factorable multivector is identified with a
certain bit vector.}

For any two determinants $|R\rangle , |S\rangle$ their exterior
product is defined as
$$|R\rangle \wedge |S\rangle = \cases
{ (-1)^{\varepsilon(R,S)} |R\cup S\rangle &if $R\cap S=\emptyset $
\cr 0 &if $R\cap S\ne \emptyset $\cr }\eqno(7)$$ where
$\varepsilon(R,S)$ is the number of pairs $(r,s)\in R\times S$
such that $r>s$. It is easy to show  that
$$\varepsilon(R,S)=|S\cap{\Delta}_{R}|\eqno(8)$$

Fock space equipped with the multiplication (7) is associative
and skew-commutative algebra that is called Grassmann or exterior
algebra of the one-electron Fock space ${\cal F}_{N,1}$.
Skew-commutativity
$$|R\rangle \wedge |S\rangle =(-1)^{|R||S|}|S\rangle \wedge
|R\rangle \eqno(9)$$ readily follows from Eqs.(2) and (8).

Creation-annihilation operators associated with spin-orbital index
$i$ are defined by the following relations
$$a_i^{\dag}|R\rangle=(-1)^{|R|}(1-\zeta_{i,R})(-1)^{|\{i\}\cap {\Delta}_{R}|}|R\cup \{i\}\rangle\eqno(10a)$$
$$a_i|R\rangle=(-1)^{|R|}\zeta_{i,R}(-1)^{|\{i\}\cap {\Delta}_{R}|}|R\backslash \{i\}\rangle\eqno(10b)$$
where
$$\zeta (I,R)=\cases {1 &if $I\subset R$\cr
                     0 &if $I\not\subset R$\cr }
\eqno(11)$$

is the well-known combinatorial $\zeta$ function of partially
ordered by inclusion set ${\cal P}(N)$\cite {Stanley}. Note that by an
abuse of notation we use symbol $\zeta (i,R)$ instead of $\zeta (\{i\},R)$.

It is not difficult to ascertain that this definition of the
creation-annihila-tion operators is identical to the commonly used
one \cite {Panin-1} .

\bigbreak \bigbreak
{\Large \bf Pointed Fock Space}

\bigbreak \bigbreak

Let us consider {\it  pointed Fock space} ${\cal A}_R=({\cal
F}_N,|R\rangle)$, that is the Fock space where some factorable
$p-$vector (determinant) $|R\rangle $ is selected. It will be
convenient to introduce the following notation for basis
determinants of the pointed Fock space:
$$e_J^I(R)=|(R\backslash J)\cup I\rangle \eqno(12)$$
It is pertinent to emphasize that Eq.(12) is just a  notation for
$2^n$ basis  determinants of the Fock space where some factorable
multivector (determinant $|R\rangle$) is selected. The selected
point $|R\rangle$ is referred to as  either the HF vacuum or the
HF reference state.

Let us define the multiplication in ${\cal A}_R$ by putting
$$e_J^I(R)\star e_{J'}^{I'}(R)= \cases{(-1)^{|(J'\cup I')\cap
{\Delta}_{(J\cup I)}|}e_{J\cup J'}^{I\cup I'}(R) &if $J\cap
J'=\emptyset$ and $I\cap I'=\emptyset$\cr 0 &if $J\cap J'\ne
\emptyset \ $ or $\ I\cap I'\ne \emptyset$\cr}.\eqno(13)$$

${\cal A}_R$ with multiplication (13) becomes associative algebra
with the identity $e_{\emptyset}^{\emptyset}(R)$. Associativity follows from
the relation
$$|(J_3\cup I_3)\cap {\Delta}_{(J_2\cup I_2})|+|(J_2\cup J_3\cup
I_2\cup I_3)\cap {\Delta}_{(J_1\cup I_1)}|$$
$$\equiv |(J_2\cup I_2)\cap {\Delta}_{(J_1\cup I_1})|+|(J_3\cup I_3)\cap
{\Delta}_{(J_1\cup J_2\cup I_1\cup I_2)}| (mod\ 2).$$

From Eq.(2) we readily obtain the equality
$$e_J^I(R)\star e_{J'}^{I'}(R)=(-1)^{|J\cup I|\cdot |J'\cup
I'|} e_{J'}^{I'}(R)\star e_J^I(R)\eqno(14)$$
which means that the multiplcation (13) is skew-commutative.

${\cal A}_R$ is a direct sum
$${\cal A}_R={\cal A}_R^{+} \oplus {\cal A}_R^{-}\eqno(15)$$
where subspaces ${\cal A}_R^{+}$ and  ${\cal A}_R^{-}$ are spanned
by basis vectors $e_J^I(R)$ with {\it even} and {\it odd} values
of $|J\cup I|$, respectively.

From Eq.(14) it follows that
$${\cal A}_R^{+}\star {\cal A}_R^{+}\subset {\cal A}_R^{+},\
{\cal A}_R^{+}\star {\cal A}_R^{-}\subset {\cal A}_R^{-},\  {\cal
A}_R^{-}\star {\cal A}_R^{+}\subset {\cal A}_R^{-},\  {\cal
A}_R^{-}\star {\cal A}_R^{-}\subset {\cal A}_R^{+}, \eqno(16)$$
which means that ${\cal A}_R$ is ${\mathbb Z}_2-$graded
(super)algebra and that ${\cal A}_R^{+}$ is a subalgebra of ${\cal
A}_R$.

Let us consider the pointed Fock space ${\cal A}_{\emptyset}$ that
is the Fock space where  the 'absolute' vacuum state $|R\rangle
=|\emptyset\rangle $  is selected. In this case we have
$$e_{\emptyset}^I(\emptyset)\star e_{\emptyset}^{I'}(\emptyset)=\cases{(-1)^{|I'\cap
{\Delta}_{I}|}e_{\emptyset}^{I\cup I'}(\emptyset) &if $I\cap
I'=\emptyset $\cr 0 &if $I\cap I'\ne \emptyset $\cr}\eqno(17)$$
and, consequently, the multiplication (13) in this case is identical
to the standard exterior product in the Fock space as follows from
Eqs.(7) and (8).

There exists another multiplication in ${\cal A}_R$ defined by the relation
$$e_J^I(R)\bullet e_{J'}^{I'}(R)= \cases{(-1)^{|(J\cup I)\cap
{\Delta}_{(J'\cup I')}|}e_{J\cup J'}^{I\cup I'}(R) &if $J\cap
J'=\emptyset$ and $I\cap I'=\emptyset$\cr 0 &if $J\cap J'\ne
\emptyset \ $ or $\ I\cap I'\ne \emptyset$\cr}.\eqno(18)$$ With
this multiplication ${\cal A}_R$ is also ${\mathbb Z}_2-$graded
(super)algebra. The pointed Fock space ${\cal A}_{\emptyset}$ with
bullet product (18) is, however, not identical to the Grassmann
algebra. It is easy to show that for arbitrary $x,y\in {\cal A}_R$
$$x\star y=y\bullet x\eqno(19)$$
that is star and bullet products are just mutually opposite
composition laws.

Let us denote by the symbol ${\cal A}_R^p$ the $p-$electron sector
of the Fock space where some determinant $|R\rangle $ is selected.
In other words, ${\cal A}_R^p$ is a pointed exterior power of the
one-electron Fock space. ${\cal A}_R^p$ is spanned by basis
elements $e_J^I(R)$ with $J\subset R, I\subset N\backslash R$, and
$|J|=|I|$. It is easy to see that both the star product (13) and
the bullet product (18) coincide on ${\cal A}_R^p$ (see Eq.(19))
and define on ${\cal A}_R^p$ the structure of associative and
commutative algebra. This structure turned out to be closely
related to the coupled cluster approach \cite {Bartlett-1}- \cite
{Crawford}.

In the remainder part of this paper we will consider pointed Fock
space as a superalgebra with the star product (13).

Let us examine the action of the creation-annihilation operators
on star product of basis determinants. We start with two formulas
that readily follows from Eqs.(10a) and (10b):
$$a_i^{\dagger}e_J^I(R)= (-1)^{\gamma}\left
[ {\zeta}_{i,J}e_{J\backslash \{i\}}^I(R)+{\zeta}_{i,N\backslash
R\backslash I} e_J^{I\cup \{i\}}(R)\right ]\eqno(20a)$$
$$a_ie_J^I(R)= (-1)^{\gamma}\left
[ {\zeta}_{i,R\backslash J}e_{J\cup \{i\}}^I(R)+{\zeta}_{i,I}
e_J^{I\backslash \{i\}}(R)\right ]\eqno(20b)$$ where
$$\gamma=|(R\backslash J)\cup I|+
|\{i\}\cap{\Delta}_{(R\backslash J)\cup I}|.\eqno(21)$$

Simple but somewhat tedious manipulations with phase prefactors lead to

$$a_i^{\dagger}\left [e_J^I(R)\star e_{J'}^{I'}(R)\right ]=
{\zeta}_{i,N\backslash R}\left [
a_i^{\dagger}e_J^I(R)\right ] \star e_{J'}^{I'}(R)$$
$$+{\zeta}_{i,R}\left \{ \left [a_i^{\dagger}e_J^I(R)\right ]
\star e_{J'}^{I'}(R)+ (-1)^{|J\cup I|}e_J^I(R)\star \left [
a_i^{\dagger}e_{J'}^{I'}(R)\right ]\right \} \eqno(22a)$$

$$a_i\left [e_J^I(R)\star e_{J'}^{I'}(R)\right ]=
{\zeta}_{i,R}\left [a_ie_J^I(R)\right ] \star e_{J'}^{I'}(R)$$
$$+{\zeta}_{i,N\backslash R}\left \{ \left [
a_ie_J^I(R)\right ] \star e_{J'}^{I'}(R) + (-1)^{|J\cup I|}
e_J^I(R)\star \left [a_ie_{J'}^{I'}(R)\right ] \right
\}\eqno(22b)$$

$$\left [ a_i^{\dagger}e_J^I(R)\right ] \star e_{J'}^{I'}(R)=
 (-1)^{|J\cup I|}e_J^I(R)\star \left
[ a_i^{\dagger}e_{J'}^{I'}(R)\right ], \ \mbox {if}\  i\in
N\backslash R\eqno(22c)$$

$$\left [ a_ie_J^I(R)\right ] \star e_{J'}^{I'}(R)=
 (-1)^{|J\cup I|}e_J^I(R)\star \left
[ a_ie_{J'}^{I'}(R)\right ], \ \mbox {if}\  i\in  R\eqno(22d)$$

Combination of Eq.(22a)-(22d) gives

$$a_i^{\dagger}\left [e_J^I(R)\star e_{J'}^{I'}(R)\right ]=$$
$$\frac{1}{1+{\zeta}_{i,N\backslash R}}\left \{ \left [a_i^{\dagger}e_J^I(R)\right ]
\star e_{J'}^{I'}(R)+ (-1)^{|J\cup I|}e_J^I(R)\star \left [
a_i^{\dagger}e_{J'}^{I'}(R)\right ]\right \}\eqno(23a)$$ and

$$a_i\left [e_J^I(R)\star e_{J'}^{I'}(R)\right ]=$$
$$\frac{1}{1+{\zeta}_{i,R}}\left \{ \left [
a_ie_J^I(R)\right ] \star e_{J'}^{I'}(R) + (-1)^{|J\cup I|}
e_J^I(R)\star \left [a_ie_{J'}^{I'}(R)\right ] \right
\}\eqno(23b)$$

\bigbreak \bigbreak

{\Large \bf Pointed p-Electron Sector of the Fock Space}

\bigbreak \bigbreak

Let  ${\cal A}_R^p$ be the $p-$electron sector ${\cal F}_{N,p}$ of
the Fock space where some determinant $|R\rangle$ is selected. As
has already been mentioned, the vector space ${\cal A}_R^p$ with
star product (13) is a commutative algebra with
$e_{\emptyset}^{\emptyset}(R)=|R\rangle $ as its identity. ${\cal
A}_R^p$ is spanned by the basis vectors $e_J^I(R)=|(R\backslash
J)\cup I\rangle$ with $|J|=|I|$.

We illustrate star multiplication in algebra ${\cal A}_R^p$ on
simple example. Let $N=\{1,2,3,4\}$, $p=2$, and $R=\{1,2\}$. There
are six basis vectors and their multiplication rules are presented
in Table I.

\begin{table}[!ht]
{\bf TABLE I}

{\bf Basis vectors multiplication table: N=\{1,2,3,4\} and
R=\{1,2\}.} \bigbreak

\begin{tabular}{|c||c|c|c|c|c|c|}
\hline
 & $e_{\emptyset}^{\emptyset}(R)$ & $e_{\{1\}}^{\{3\}}(R)$ & $e_{\{1\}}^{\{4\}}(R)$ & $e_{\{2\}}^{\{3\}}(R)$ & $e_{\{2\}}^{\{4\}}(R)$ & $e_{\{1,2\}}^{\{3,4\}}(R)$ \\
\hline \hline
$e_{\emptyset}^{\emptyset}(R)$ & $e_{\emptyset}^{\emptyset}(R)$ & $e_{\{1\}}^{\{3\}}(R)$ & $e_{\{1\}}^{\{4\}}(R)$ & $e_{\{2\}}^{\{3\}}(R)$ & $e_{\{2\}}^{\{4\}}(R)$ & $e_{\{1,2\}}^{\{3,4\}}(R)$ \\
\hline
$e_{\{1\}}^{\{3\}}(R)$ & $e_{\{1\}}^{\{3\}}(R)$ & 0 & 0 & 0 & $-e_{\{1,2\}}^{\{3,4\}}(R)$ & 0 \\
\hline
$e_{\{1\}}^{\{4\}}(R)$ & $e_{\{1\}}^{\{4\}}(R)$ & 0 & 0 & $e_{\{1,2\}}^{\{3,4\}}(R)$ & 0 & 0 \\
\hline
$e_{\{2\}}^{\{3\}}(R)$ & $e_{\{2\}}^{\{3\}}(R)$ & 0 & $e_{\{1,2\}}^{\{3,4\}}(R)$ & 0 & 0 & 0 \\
\hline
$e_{\{2\}}^{\{4\}}(R)$ & $e_{\{2\}}^{\{4\}}(R)$ & $-e_{\{1,2\}}^{\{3,4\}}(R)$  & 0 & 0 & 0 & 0 \\
\hline
$e_{\{1,2\}}^{\{3,4\}}(R)$ & $e_{\{1,2\}}^{\{3,4\}}(R)$ & 0 & 0 & 0 & 0 & 0 \\
\hline
\end{tabular}

\end{table}

From Eqs.(23a)-(23b)) it follows that for arbitrary vectors $x\in
{\cal A}_R^p$ and  $y\in {\cal A}_R$
$$a_i^{\dagger} \left [x\star y\right ]= \frac{1}{1+{\zeta}_{i,N\backslash
R}}\left \{ (a_i^{\dagger}x)\star y+x\star  (a_i^{\dagger}y )
\right \}\eqno(24a)$$ and
$$a_i \left [x\star y\right ]= \frac{1}{1+{\zeta}_{i,R}}
\left \{ (a_i x)\star y+x\star (a_i y) \right \}\eqno(24b)$$

In view of  Eqs.(22c) and (22d) we have
$$\ \ \ \  (a_i^{\dagger}x)\star y=x\star (a_i^{\dagger}y) \ \ \mbox{if }
\ \ i\in N\backslash R\eqno(25a)$$ and
$$(a_ix)\star y=x\star (a_iy)\ \ \mbox{if} \ \ \  i\in R\eqno(25b)$$
It is pertinent to emphasize that Eqs.(24)-(25) are valid for
arbitrary $y\in {\cal A}_R$ not necessarily belonging to the
$p-$electron sector of the Fock space.

The following relations are readily obtained from Eqs.(24a)-(24b):
$$a_i^{\dagger}x^{\mu}=\frac{\mu}{1+(\mu-1){\zeta}_{i,N\backslash
R}}\left [x^{\mu-1}\star a_i^{\dagger} x\right ]\eqno(26a)$$
$$a_ix^{\mu}=\frac{\mu}{1+(\mu-1){\zeta}_{i,R}
}\left [x^{\mu-1}\star a_i x\right ]\eqno(26b)$$ where $x$ is
arbitrary vector from ${\cal A}_R^p$. Here $x^{\mu}=\underbrace
{x\star \ldots \star x}_{\mu}$.

${\cal A}_R^p$ is a Hermitean space with scalar product
$$\langle x | y\rangle =\sum\limits_{\mu=0}^p\sum\limits_{J\subset R \atop {I\subset N\backslash R}}^{(\mu)}
(x_J^I)^*y_J^I\eqno(27a)$$ and norm
$$\|x\|=\langle x|x\rangle
^{\frac{1}{2}}\eqno(27b)$$

There exists an important question concerning interrelation of
algebra structure on ${\cal A}_R^p$ and its structure as a normed
space with the norm (27b). This question may be formulated as: Is
the multiplication (13) consistent with the norm (27b), or, in
other words, does the inequality
$$\|x\star y\|\le \|x\|\cdot \|y\|\eqno(28)$$
hold true for arbitrary $x,y\in {\cal A}_R^p$? The following
simple example shows, that in general the answer to this question
is negative. If $n=2$ and $p=1$, then for
$x=e_{\emptyset}^{\emptyset}+e_{\{1\}}^{\{2\}}$ the inequality
(28) is violated. Indeed, in this case $x\star
x=e_{\emptyset}^{\emptyset}+2e_{\{1\}}^{\{2\}}$ and $\|x\star
x\|=\sqrt{5}$ whereas $\|x\|^2=2$. Computer experiments show that
for $n\le 8$ the inequality (28) does not hold true, at least if
$p=1,n-1$. Till now no violation of this inequality has been
detected for $n>8$. We may suggest that for sufficiently large $n$
the inequality (28) holds true for all $x,y\in {\cal A}_R^p$, or,
in other words, that there exists $n_0$ such that for $n>n_0$
${\cal A}_R^p$ is a unital Banach algebra. Verification of this
suggestion may be a rather complicated task.

For each pair $(k,l)$ with $k\le l$ let us introduce subspace
${\cal W}_R^{(k,l)}$ spanned by vectors $e_J^I(R)$ with $k\le
|J|=|I|\le l$. It is clear that
$${\cal A}_R^p=\bigoplus\limits_{k=0}^p {\cal W}_R^{(k,k)}\eqno(29)$$
and that
$${\cal W}_R^{(k_1,k_1)}\star {\cal
W}_R^{(k_2,k_2)}\subset {\cal W}_R^{(k_1+k_2,k_1+k_2)}\eqno(30)$$
where $k_1+k_2\le p$, which means that ${\cal A}_R^p$ is a graded
algebra. The subspace ${\cal W}_R^{(1,p)}$ is its maximal
nilpotent ideal and the algebra under discussion is just a direct
sum of the field of complex numbers and this nilpotent ideal
$${\cal A}_R^p=\mathbb C e_{\emptyset}^{\emptyset}(R)\oplus {\cal
W}_R^{(1,p)}\eqno(31)$$ Note also that ${\cal A}_R^p$ is algebra
with involution induced by the complex conjugation.

Spectral theory (see, e.g., \cite {Bourbaki}) in algebra ${\cal
A}_R^p$ is very simple. Indeed, let us write  arbitrary element $x
\in {\cal A}_R^p$ as
$$x=x_{\emptyset}^{\emptyset}e_{\emptyset}^{\emptyset}+\tau\eqno(32)$$
 where $\tau \in {\cal W}_R^{(1,p)}.$
Its spectrum is defined as the set of all $\lambda \in \mathbb C$
such that $(\lambda e_{\emptyset}^{\emptyset}-x)$ is not
invertible in ${\cal A}_R^p$. It is easy to see that $$(\lambda
e_{\emptyset}^{\emptyset}-x)^{-1}=\sum\limits_{k=0}^p\frac{{\tau}^k}{(\lambda
- x_{\emptyset}^{\emptyset})^{k+1}}={\cal R}(x,\lambda)\eqno(33)$$
exists for all $\lambda \ne x_{\emptyset}^{\emptyset}$. Thus, any
element $x\in {\cal A}_R^p$ has a single point spectrum

$$Sp\ x=\{\langle x|e_{\emptyset}^{\emptyset}\rangle \}\eqno(34)$$
and  resolvent   $\mathbb C\backslash
\{x_{\emptyset}^{\emptyset}\}\to {\cal A}_R^p $ defined by
Eq.(33). In particular, $Sp \ \tau=\{0\}$ for any element $\tau
\in {\cal W}_R^{(1,p)}$.

 It is well-known  that in commutative algebras of
the type of Eq.(31) it is easy to define {\it algebraically} both
the exponential mapping and its inverse. Namely,
$$exp:\tau\to \sum\limits_{{\mu}=0}^p\frac{{\tau}^{\mu}}{{\mu}!}\eqno(35)$$
is the exponential mapping ${\cal W}_R^{(1,p)}\to
e_{\emptyset}^{\emptyset}(R)+{\cal W}_R^{(1,p)}$ and
$$log: e_{\emptyset}^{\emptyset}(R)+\tau\to \sum\limits_{\mu=1}^p
(-1)^{\mu-1}\frac{{\tau}^{\mu}}{\mu}\eqno(36)$$ is the logarithmic
mapping $e_{\emptyset}^{\emptyset}(R)+{\cal W}_R^{(1,p)}\to {\cal
W}_R^{(1,p)}$.

For mappings (35) and (36) all classic relations hold true:
$$exp({\tau}_1)\star exp({\tau}_2)=exp({\tau}_1+{\tau}_2),\eqno(37a)$$
$$[exp(\tau)]^{-1}=exp(-\tau),\eqno(37b)$$
$$exp(log(e_{\emptyset}^{\emptyset}(R)+\tau))=\tau,\eqno(37c)$$ and
$$log(exp(\tau))=\tau,\eqno(37d)$$ where $\tau, {\tau}_1, {\tau}_2$
are arbitrary elements from ${\cal W}_R^{(1,p)}$.

Using Eqs.(24)-(26), it is possible to calculate explicitly the
action of particle number conserving products of the
creation-annihilation operators on powers of state vectors
(elements of algebra ${\cal A}_R^p$). For  the simplest product
$a_i^{\dagger}a_j$ we have
$$a_i^{\dagger}a_jx^{\mu}=
\cases { -(\mu -1){\delta}_{ij}x^{\mu}+\mu x^{\mu -1}\star
a_i^{\dagger}a_jx &if $i,j\in R$\cr \mu x^{\mu-2}\star \left
[(\mu-1)a_i^{\dagger}x\star a_jx+x\star a_i^{\dagger}a_j x\right ]
&if $i\in R,j\in N\backslash R$\cr x^{\mu -1}\star
a_i^{\dagger}a_jx &if $i\in N\backslash R,j\in R$\cr \mu x^{\mu
-1}\star a_i^{\dagger}a_jx &if $i,j\in N\backslash R$\cr
}\eqno(38)$$ Thus, the action of the operator $a_i^{\dagger}a_j$
on arbitrary power of vector $x\in {\cal A}_R^p$ is completely
determined by the action of three operators $a_i^{\dagger}$,
$a_j$, and $a_i^{\dagger}a_j$ on this vector. It is pertinent to
note that the expression $a_i^{\dagger}x\star a_jx$ involved in
the right-hand side of Eq.(38) is not reduced to $x\star
a_i^{\dagger}a_jx$ (see restrictions on indices in
Eqs.(25a)-(25b)).

Many methods of quantum chemistry are based on simple idea of
parame-trization of many electron wave functions with a relatively
small number of free parameters with their subsequent optimization
using one or other optimality criterium. In particular, in CC and
related approaches  subsets of ${\cal A}_R^p$ are parametrized
with the aid of  elements ('amplitudes') from ${\cal W}_R^{(1,l)}$
where $l$ is the maximal CC excitation level. General
parametrization of such a type may be defined as a differentiable
mapping
$$\pi:{\cal W}_R^{(1,l)}\to {\cal A}_R^p\eqno(39)$$
satisfying two requirements:

(i) $\pi$ is an injective mapping;

(ii) $\pi(\tau)$ is invertible for any $\tau \in {\cal
W}_R^{(1,l)}$.

 The last requirement guarantees that $\langle
\pi (\tau)|e_{\emptyset}^{\emptyset}\rangle \ne 0 $, or, in other
words, that the HF vacuum state $|R\rangle $ appears with nonzero
coefficient in expansion of $\pi(\tau)$ for each $\tau$.

Polynomial mapping
$$P_a:\tau\to \sum\limits_{\mu=0}^pa_{\mu}{\tau}^{\mu},\eqno(40)$$
where $a=(a_0,a_1,\ldots,a_p)\in {\mathbb C}^{p+1}$,  is a
parametrization if and only if $a_0\cdot a_1\ne 0$. Indeed,
$a_0\ne 0$ means that $P_a(\tau)$ is invertible for each $\tau$.
The mapping inverse to $P_a$ may be written in the form
$$P_a^{-1}:a_0e_{\emptyset}^{\emptyset}+\tau \to
\sum\limits_{\mu=1}^p(-1)^{\mu-1}b_{\mu}{\tau}^{\mu}, \eqno(41)$$
where coefficients $b_{\mu}$ are easily obtained from the
recurrence relation
$$b_{\mu}=\frac{(-1)^{\mu}}{a_1^{\mu}}\sum\limits_{\nu=1}^{\mu-1}(-1)^{\nu-1}b_{\nu}
\sum\limits_{i_1+\cdots +i_{\nu}=\mu}a_{i_1}\cdots
a_{i_{\nu}},\eqno(42a)$$
$$ b_1=\frac{1}{a_1}.\eqno(42b)$$
For example, QCI parametrization $\tau \to {\cal R}(\tau,1)$
corresponds to the case $a_0=a_1=\cdots=a_p=1$ and from
Eqs.(42a)-(42b) we easily obtain $b_1=\cdots=b_p=1$. For the
exponential parametrization these equations give  $b_{\mu}=\frac
{1}{\mu}$ in full accordance with Eq.(35).

It is clear that without loss of generality we can put
$a_0=a_1=1$. Indeed, the condition $a_0=1$ means that the
intermediate normalization condition $\langle
x|e_{\emptyset}^{\emptyset}\rangle=1$ for state vectors is used.
Going from arbitrary nonzero $a_1$ to $a_1=1$ simply corresponds
to scaling of vector $\tau$ of CC amplitudes and such scaling is
of no consequence if components of $\tau$ vector and coefficients
$a_1,a_2,\ldots,a_p$  are considered as free parameters.

 There exits as well standard integral representation for
polynomial functions on ${\cal W}_R^{(1,p)}$:
$$P_a(\tau)= \frac {1}{2\pi
i}\oint\limits_{|\lambda|=1}P_a(\lambda){\cal
R}(\tau,\lambda)d\lambda\eqno(43)$$ where $\lambda$ is complex
variable. This integral representation, being trivial for
polynomials, could be extended to certain algebras of analytic
functions if our hypothesis that ${\cal A}_R^p$ is a Banach
algebra, turns out to be true.

 Let us suppose that polynomial parametrization covers
some eigenvector of the electronic Hamiltonian $H$, that is there
exist coefficients $a=(1,1,a_2,\ldots,a_p)$ and vector $\tau \in
{\cal W}_R^{(1,l)}$ such that
$$HP_a(\tau)=EP_a(\tau).\eqno(44)$$
Let us put
$$P_{a,E}:\tau \to EP_a(\tau) \eqno(45)$$
Its inverse is
$$P_{a,E}^{-1}:Ee_{\emptyset}^{\emptyset}+\tau \to
\sum\limits_{\mu=1}^p(-1)^{\mu-1}\frac{b_{\mu}}{E^{\mu}}{\tau}^{\mu}\eqno(46)$$
Application of this transformation to the both sides of Eq.(44)
gives
$$\sum\limits_{\mu=1}^p(-1)^{\mu -1}\frac{b_{\mu}}{E^{\mu}}\left
[HP_a(\tau)-Ee_{\emptyset}^{\emptyset}\right
]^{\mu}=\tau\eqno(47)$$

This relation may serve as a base for iterative algorithms for
solution of Eq.(44).

The {\it polynomial} derivative of $P_a(\tau)$ in algebra ${\cal
A}_R^p$ is defined as
$$\frac{d}{d\tau}P_a(\tau)=\sum\limits_{k>0}ka_k{\tau}^{k-1}\eqno(48)$$
Partial derivatives of $P_a(\tau)$ with respect to components of
vector $\tau$ are connected with its polynomial derivative by the
relation
$$\frac{\partial}{\partial
t_J^I}P_a(\tau)=\left [\frac{d}{d\tau}P_a(\tau)\right ]\star
e_J^I(R)\eqno(49)$$ Note that ${\tau}^p$, if nonzero, is a linear
combination of basis vectors $e_R^{I'}(R)$ where ${I'}\subset
N\backslash R$ with  $|I'|=p$. As a result, ${\tau}^p\star
e_J^I(R)=0$ for any basis vector $e_J^I(R)$  with $J\ne \emptyset$
and
$$\frac{\partial}{\partial
t_J^I}exp(\tau)=exp(\tau)\star e_J^I(R)\eqno(50)$$

 If $\tau \in {\cal W}_R^{(1,l)}$ is a vector of
CC amplitudes then the coefficients in expansion of $exp(\tau)$ in
algebra ${\cal A}_R^p$ with respect to the basis $e_J^I(R)$ are
just the CC configuration interaction (CI) coefficients:
$$exp(\tau)=\sum\limits_{\mu=0}^p\sum\limits_{J\subset R\atop{I\subset N\backslash
R}}^{(\mu)}T_{\mu}^l(J,I)e_J^I(R),\eqno(51a)$$ where
$$T_{\mu}^l(J,I)=\langle exp(\tau)|e_J^I(R)\rangle . \eqno(51b)$$

 {\it Thus, $p$-electron sector of the Fock space equipped with the structure of
commutative algebra, defined by Eq.(13), may be considered as a
natural domain for the CC and related approaches.}

With the aid of Eq.(26b) it is easy to get an expression for
matrix elements
$${\rho}_{ij}=\langle exp(\tau)|a_i^{\dagger}a_j|exp(\tau)\rangle=\langle a_iexp(\tau)|
a_jexp(\tau)\rangle\eqno(52)$$ of (not normalized) CC 1-density
operator. We have
$${\rho}_{ij}=\cases{
\langle s_1(\tau)\star a_i\tau|s_1(\tau)\star a_j\tau\rangle &if
$i,j\in R$\cr \langle s_1(\tau)\star a_i\tau|s_2(\tau)\star
a_j\tau\rangle &if $i\in R,j\in N\backslash R$\cr \langle
s_2(\tau)\star a_i\tau|s_1(\tau)\star a_j\tau\rangle &if $i\in
N\backslash R,j\in R$\cr \langle s_2(\tau)\star
a_i\tau|s_2(\tau)\star a_j\tau\rangle &if $i,j\in N\backslash
R$\cr }\eqno(53)$$ where
$$s_1(\tau)=\sum\limits_{\mu=0}^{p-1}\frac{{\tau}^{\mu}}{{(\mu}+1)!}\eqno(54a)$$
and
$$s_2(\tau)=\frac{d}{d\tau}exp(\tau)\eqno(54b)$$

Note that the simplicity of Eq.(53) is somewhat misleading because
in this equation two sectors of the Fock space are involved. If it
is desirable to stay in $p-$electron sector then one should employ
Eq.(38) with direct expansion of the product $a_i^{\dagger}x\star
a_jx$ via basis $p-$electron vectors $e_J^I(R)$.

\bigbreak

{\Large \bf  Non-Standard Realizations of p-Electron }

{\Large \bf Sector  of the Fock Space} \bigbreak

 In this section we
consider realizations of the $p-$electron sector of the Fock space
suggested in \cite {Panin-2,Panin-3}. The main idea consists in
replacing $p-$electron state vector by a certain family of
$q-$electron vectors. If $p+q=n$ then there exists a well-known
isomorphism between $q-$ and $p-$electron sectors of the Fock
space that is called Hodge isomorphism and is denoted by the
symbol $*$\cite {Hodge}:
$$*({\psi}_{i_1}\wedge \ldots \wedge {\psi}_{i_q})
=(-1)^{|J\cap {\Delta}_N|}{\psi}_{j_1}\wedge \ldots \wedge
{\psi}_{j_p}\eqno(55)$$ where $I=i_1<\ldots <i_q$, $J=j_1<\ldots
<j_p$, and $I\cup J=N$.  Eq.(55)  defines linear mapping
$$*:\bigwedge^q {\cal F}_{N,1}\to \bigwedge^{n-q} {\cal F}_{N,1}.\eqno(56)$$
that does not depend on the choice of the (orthonormal) MSO basis
set in the case when the one-electron sector of the Fock space is
Euclidean and oriented one. For complex Fock space Eq.(55)
certainly defines isomorphic mapping  but this mapping may depend
on the choice of MSO basis set.  Note that in Eq.(55) we
introduced the phase prefactor that differs from that in \cite
{Hodge} by the constant factor $(-1)^{[\frac{p+1}{2}]}$. Note as
well that the Hodge isomorphism is not an involution.

In general case we can consider a family $\{x_q^Z\}_{Z\subset N}$
of $q-$electron vectors indexed by $(p+q)-$element subsets of the
index set $N$ where $x_q^Z\in {\cal F}_{Z,q}$ and apply Hodge
isomorphism $*:{\cal F}_{Z,q}\to {\cal F}_{Z,p}$ to each vector
$x_q^Z$ (by an abuse of notation we use the same symbol for Hodge
isomorphisms defined on different subspaces) :
$$*\left (\sum\limits_{S\subset Z}^{(q)}\langle S|x_q^Z\rangle|S\rangle \right )=
\sum\limits_{S\subset Z}^{(q)}(-1)^{|(Z\backslash
S)\cap{\Delta}_Z|} \langle S|x_q^Z\rangle|Z\backslash S\rangle
\eqno(57)$$ If the family $\{x_q^Z\}_{Z\subset N}$ satisfies the
conditions
$$(-1)^{|(Z\backslash S)\cap{\Delta}_Z|}\langle S|x_q^Z\rangle =
(-1)^{|(Z'\backslash S')\cap{\Delta}_{Z'}|}\langle
S'|x_q^{Z'}\rangle \eqno(58)$$ for all pairs $(Z,S),(Z',S')$ such
that
$$S\subset Z,\ S'\subset Z', \ \mbox{and} \ Z\backslash S = Z'\backslash S'.\eqno(59)$$
then it is possible to assemble {\it the unique} $p-$electron
vector
$$x_p=\frac{1}{{{n-p}\choose {q}}}\sum\limits_{Z\subset N}^{(p+q)}*\left (x_q^Z\right )\eqno(60)$$

Each family $\{x_q^Z\}_{Z\subset N}$ of $q-$electron vectors
satisfying the conditions (58) will be called $(p,q)-$vector.
Members $x_q^Z$ of this family will be called its $q-$components.
For the set of all $(p,q)-$vectors the symbol ${\cal S}_{N,p,q}$
will be used. The mapping
$$s_{q\uparrow p}:\left \{x_q^Z\right \}_{Z\subset N}\to \{*\left (x_q^Z\right )\}_{Z\subset N}
\to \frac{1}{{{n-p}\choose {q}}} \sum\limits_{Z\subset
N}^{(p+q)}*\left (x_q^Z\right )\eqno(61)$$ of the set ${\cal
S}_{N,p,q}$ to the $p-$electron sector of the Fock space is called
'the assembling mapping'. It is easy to ascertain that it is a
bijection and its inverse is the so-called disassembling mapping
defined as
$$s_{p\downarrow q}:x_p\to \left \{ \sum\limits_{R\subset
Z}^{(p)}(-1)^{|R\cap{\Delta}_Z|}\langle R|x_p\rangle |Z\backslash
R\rangle \right \}_{Z\subset N} \eqno(62)$$

Let us consider simple example. For $N=\{1,2,3,4\}$, $p=2$, and
$q=1$ there are four $3-$element subsets:
$$Z_1=\{1,2,3\},\ Z_2=\{1,2,4\},\ Z_3=\{1,3,4\},\ Z_4=\{2,3,4\}.$$The
corresponding subsets ${\Delta }_{Z_i}$ are:
$${\Delta}_{Z_1}=\{1,3\},\ {\Delta}_{Z_2}=\{1,3,4\},\ {\Delta}_{Z_3}=\{1,4\},\ {\Delta}_{Z_4}=\{1,2,4\}$$
Arbitrary family $\{x_1^{Z_i}\}_{i=1}^4 $ is constituted by the
vectors
$$x_1^{Z_1}=c_1^{123}|1\rangle + c_2^{123}|2\rangle +
c_3^{123}|3\rangle \in {\cal F}_{Z_1,1}$$
$$x_1^{Z_2}=c_1^{124}|1\rangle + c_2^{124}|2\rangle +
c_4^{124}|4\rangle\in {\cal F}_{Z_2,1}$$
$$x_1^{Z_3}=c_1^{134}|1\rangle + c_3^{134}|3\rangle +
c_4^{134}|4\rangle\in {\cal F}_{Z_3,1}$$
$$x_1^{Z_4}=c_2^{234}|2\rangle + c_3^{234}|3\rangle +
c_4^{234}|4\rangle\in {\cal F}_{Z_4,1}$$ In the case under
consideration there are six relations of the type of Eq.(59):
$$Z_1\backslash \{1\}=Z_4\backslash \{4\}=\{2,3\},\ Z_1\backslash
\{2\}=Z_3\backslash\{4\}=\{1,3\},$$
$$Z_1\backslash \{3\}=Z_2\backslash \{4\}=\{1,2\},\ Z_2\backslash
\{1\}=Z_4\backslash\{3\}=\{2,4\},$$
$$Z_2\backslash \{2\}=Z_3\backslash \{3\}=\{1,4\},\ Z_3\backslash
\{1\}=Z_4\backslash\{2\}=\{3,4\}.$$ Using Eq.(58), we obtain
$$c_1^{123}=c_4^{234}=C_{23},\ c_2^{123}=-c_4^{134}=C_{13},$$
$$c_3^{123}=c_4^{124}=C_{12},\ c_1^{124}=-c_3^{234}=C_{24},$$
$$c_2^{124}=c_3^{134}=C_{14},\ c_1^{134}=c_2^{234}=C_{34},$$
and family $\{x_1^{Z_i}\}_{i=1}^4 $ is $(2,1)-$vector if and only
if
$$x_1^{Z_1}=C_{23}|1\rangle + C_{13}|2\rangle +
C_{12}|3\rangle $$
$$x_1^{Z_2}=C_{24}|1\rangle + C_{14}|2\rangle +
C_{12}|4\rangle$$
$$x_1^{Z_3}=C_{34}|1\rangle + C_{14}|3\rangle -
C_{13}|4\rangle$$
$$x_1^{Z_4}=C_{34}|2\rangle - C_{24}|3\rangle +
C_{23}|4\rangle $$ Application of  Hodge isomorphism to each
1-component of this vector gives
$$*\left (x_1^{Z_1}\right )=-C_{23}|23\rangle + C_{13}|13\rangle -
C_{12}|12\rangle $$
$$*\left (x_1^{Z_2}\right )=-C_{24}|24\rangle + C_{14}|14\rangle -
C_{12}|12\rangle$$
$$*\left (x_1^{Z_3}\right )=-C_{34}|34\rangle + C_{14}|14\rangle +
C_{13}|13\rangle$$
$$*\left (x_1^{Z_4}\right )=-C_{34}|34\rangle - C_{24}|24\rangle -
C_{23}|23\rangle $$ The corresponding uniquely determined
$2-$electron vector is
$$\frac{1}{2}\sum\limits_{i=1}^4*\left (x_1^{Z_i}\right
)=$$ $$=-C_{12}|12\rangle + C_{13}|13\rangle + C_{14}|14\rangle
-C_{23}|23\rangle -C_{24}|24\rangle -C_{34}|34\rangle $$

Let us put
$$\lambda \left \{x_q^Z\right \}_{Z\subset N}+\mu \left \{y_q^Z\right \}_{Z\subset N} =
\left \{\lambda x_q^Z+\mu y_q^Z\right \}_{Z\subset N}\eqno(63)$$
where $\lambda ,\mu \in {\mathbb C}$.

Due to linearity of conditions (58) the right-hand side of Eq.(63)
is a $(p,q)-$vector and this equation defines vector space
structure on the set ${\cal S}_{N,p,q}$. It is clear that the
disassembling mapping is an isomorphism of vector spaces ${\cal
F}_{N,p}$ and ${\cal S}_{N,p,q}$.

Scalar product in ${\cal S}_{N,p,q}$ consistent with that in
${\cal F}_{N,p}$ is
$$\left \langle \left \{x_q^Z\right \}_{Z\subset N}|\left \{y_q^Z\right \}_{Z\subset N}\right \rangle =
\frac{1}{{{n-p}\choose {q}}}\sum\limits_{Z\subset N}^{(p+q)} \left
\langle x_q^Z|y_q^Z\right \rangle \eqno(64)$$

Thus, for fixed $N$ and $p$ with vector space ${\cal F}_{N,p}$ it
is possible to associate its (isomorphic) models ${\cal
S}_{N,p,1},{\cal S}_{N,p,2},\ldots $.

For each $(p+q)-$element subset $Z\subset N$ let us introduce the
mapping
$${\pi}_{p,q}^Z:\left \{x_q^{Z'}\right \}_{Z'\subset N}\to x_q^Z\eqno(65)$$
that is obviously a surjective homomorphism (projection) of ${\cal
S}_{N,p,q}$ on ${\cal F}_{Z,q}$. Application to $q-$electron
vector $x_q^Z$ the Hodge isomorphism with subsequent disassembling
of vector $*(x_q^Z)$ gives
$$s_{p\downarrow q}*(x_q^Z)=\left \{ \sum\limits_{R\subset Z'\cap Z}^{(p)}(-1)^{|
R\cap{\Delta}_Z|+|R\cap{\Delta}_{Z'}|}\langle Z\backslash
R|x_q^Z\rangle |Z'\backslash R\rangle \right \}_{Z'\subset
N}\eqno(66)$$ From this equation it readily follows that
$${\pi}_{p,q}^Z s_{p\downarrow q}*\left (x_q^Z\right )=x_q^Z\eqno(67)$$
or, in other words, the mapping $s_{p\downarrow q} *$ is a right
inverse of the projection ${\pi}_{p,q}^Z$, being  linear injective
mapping. Note that for obvious reason we omit the composition
symbol in writing composition of mappings.

Eq.(67) means that any $(p,q)-$vector of the type of Eq.(66)
involves initial $q-$electron vector $x_q^Z$ as its $q-$component.
We say that $(p,q)-$vector (66) is generated by its $q-$component
$x_q^Z$. Of course, not each $(p,q)-$vector is generated by some
of its $q-$component but those, that are generated, are so
important that deserve special name.

{\bf Definition.} $(p,q)-$vector is called simple if it is
generated by one of its $q-$components, that is if there exists
$(p+q)-$element subset $Z\subset N$ such that the following
equality holds true:
$$s_{p\downarrow q}*{\pi}_{p,q}^Z \left (\left \{x_q^{Z'}\right \}_{Z'\subset N} \right
)=\left \{x_q^{Z'}\right \}_{Z'\subset N}\eqno(68)$$

Returning to the example, discussed above, we have
$$s_{2\downarrow 1}*\left (x_1^{Z_1}\right )= \left( \begin{array}{c}
{c_1^{123}|1\rangle + c_2^{123}|2\rangle + c_3^{123}|3\rangle }\\
{c_3^{123}|4\rangle}\\
{-c_2^{123}|4\rangle}\\
{c_1^{123}|4\rangle}
 \end{array}\right)$$
Here $(2,1)-$vector is represented as a column vector with entries
indexed by subsets $Z_1,Z_2,Z_3$, and $Z_4$.

 Each non-zero
$q-$component $x_q^Z$ of a given $(p,q)-$vector $\left \{ x_q^Z
\right \}_{Z\subset N}$ generates some simple $(p,q)-$vector
$s_{p\downarrow q}*(x_q^Z)$ and it is easy to ascertain that the
initial $(p,q)-$vector belongs to the linear hull $V\left (\left
\{ x_q^Z \right \}_{Z\subset N}\right )$ of these simple
$(p,q)-$vectors:
$$\left \{ x_q^Z \right \}_{Z\subset N}\in V\left (\left \{ x_q^Z \right \}_{Z\subset N}\right )
=\sum\limits_{Z\subset N}^{(p+q)}{\mathbb C}s_{p\downarrow
q}*{\pi}_{p,q}^Z\left (\left \{ x_q^Z \right \}_{Z\subset N}\right
)\eqno(69)$$

Since Hodge isomorphism in the case of Euclidean Fock space does
not depend upon the choice of (orthonormal) MSO basis, the
correspondence
$$x_p\to V\left (s_{p\downarrow q}(x_p)\right )\eqno(70)$$
supplies us with a family of invariants associated with a given
{\it real} $p-$electron vector $x_p$.

There exist obvious isomorphisms between spaces with different
$q$:
$${\cal S}_{N,p,q}\stackrel {s_{q\uparrow p}}{\longrightarrow}{\cal F}_{N,p}
\stackrel {s_{p\downarrow q'}}{\longrightarrow}{\cal
S}_{N,p,q'}\eqno(71)$$

Disassembling of arbitrary factorable $p-$vector (determinant)
$|R\rangle $ gives
$$s_{p\downarrow q}(|R\rangle )=\left \{(-1)^{|R\cap {\Delta}_Z|}|Z\backslash
R\rangle \right \}_{Z\supset R}\eqno(72)$$

It is clear that any nonzero $q-$component of the vector (72)
generates it. It is easy to prove that it is a characteristic
property of disassembled factorable $p-$vectors: $(p,q)-${\it
vector corresponds to some factorable $p-$vector (determinant) if
and only if it is generated by any of its nonzero $q-$components.}

Direct use of Eq.(64) shows that simple $(p,q)-$vectors
$s_{p\downarrow q}(|R\rangle )$ constitute orthonormal (with
respect to scalar product (64)) basis of the vector space ${\cal
S}_{N,p,q}$ of all $(p,q)-$vectors.

Now let us consider the pointed space ${\cal S}_R^{p,q}=({\cal
S}_{N,p,q},s_{p\downarrow q}(|R\rangle) )$. It is easy to endow it
with commutative algebra structure by direct transfer of the
corresponding structure from ${\cal A}_R^p$:
$$ \left \{ x_q^Z \right \}_{Z\subset N}\star \left \{y_q^Z\right \}_{Z\subset N}=
s_{p\downarrow q}\left ( s_{q\uparrow p}\left (\left \{x_q^Z\right
\}_{Z\subset N}\right )\star s_{q\uparrow p}\left (\left
\{y_q^Z\right \}_{Z\subset N}\right )\right ) \eqno(73)$$ It would
be desirable, however, to give a definition for multiplication of
$(p,q)-$vectors that does not appeal explicitly to the assembling
mapping. It is easy to see that such multiplication can not be
componentwise.

First let us select convenient basis in each $q-$electron subspace
${\cal F }_{Z,q}$. To this end we consider pairs of subsets
$J\subset R$ and $I\subset N\backslash R$ such that
  $$R\backslash (R\cap Z)\subset J\subset R ;\eqno(74a)$$
  $$I\subset Z\backslash (R\cap Z);\eqno(74b)$$
  $$|J|=|I|.\eqno(74c)$$
The total number of such pairs is equal to
$$\sum\limits_{k\ge 0}{{|R\cap Z|}\choose {k}}{{p+q-|R\cap Z|}\choose {|R\backslash (R\cap
Z)|+k}}={{p+q}\choose {p}}=dim \ {\cal F }_{Z,q}\eqno(75)$$ Basis
vectors in ${\cal F }_{Z,q}$ may be written as
$$e_J^I(Z,R)=(-1)^{|((R\backslash J)\cup I)\cap {\Delta}_Z|}e_{R\backslash (Z\cap J)}^{Z\backslash R\backslash
I}(R)\eqno(76)$$ Selection of this concrete form of basis vectors
is motivated by the relation
$$s_{p\downarrow q}(e_J^I(R))=\left \{
e_J^I(Z,R)\right \}_{Z\supset (R\backslash J)\cup I}\eqno(77)$$
that holds true for arbitrary basis determinant $e_J^I(R)$.

  Let us return to the example discussed above and
suppose that we selected $R=\{1,2\}$. Then

(1) In the vector space ${\cal F}_{Z_1,1}$ basis vectors indexed by subsets
$$J\subset R,\ I\subset \{3\}$$
are
$$e_{\emptyset}^{\emptyset}(Z_1,R)=-e_{R}^{\{3\}}(R)=-|3\rangle$$
$$e_{\{1\}}^{\{3\}}(Z_1,R)=-e_{\{2\}}^{\emptyset }(R)=-|1\rangle$$
$$e_{\{2\}}^{\{3\}}(Z_1,R)=e_{\{1\}}^{\emptyset }(R)=|2\rangle$$
(2) In the vector space ${\cal F}_{Z_2,1}$ basis vectors indexed by subsets
$$J\subset R,\ I\subset \{4\}$$
are
$$e_{\emptyset}^{\emptyset}(Z_2,R)=-e_{R}^{\{4\}}(R)=-|4\rangle$$
$$e_{\{1\}}^{\{4\}}(Z_2,R)=-e_{\{2\}}^{\emptyset }(R)=-|1\rangle$$
$$e_{\{2\}}^{\{4\}}(Z_2,R)=e_{\{1\}}^{\emptyset }(R)=|2\rangle$$
(3) In the vector space ${\cal F}_{Z_3,1}$ basis vectors indexed by subsets
$$\{2\}\subset J\subset R,\ I\subset \{3,4\}$$
are
$$e_{\{2\}}^{\{3\}}(Z_3,R)=-e_{R}^{\{4\}}(R)=-|4\rangle$$
$$e_{\{2\}}^{\{4\}}(Z_3,R)=e_{R}^{\{3\} }(R)=|3\rangle$$
$$e_{\{1,2\}}^{\{3,4\}}(Z_3,R)=-e_{\{2\}}^{\emptyset }(R)=-|1\rangle$$
(4) In the vector space ${\cal F}_{Z_4,1}$ basis vectors indexed by subsets
$$\{1\}\subset J\subset R,\ I\subset \{3,4\}$$
are
$$e_{\{1\}}^{\{3\}}(Z_4,R)=-e_{R}^{\{4\}}(R)=-|4\rangle$$
$$e_{\{1\}}^{\{4\}}(Z_4,R)=e_{R}^{\{3\} }(R)=|3\rangle$$
$$e_{\{1,2\}}^{\{3,4\}}(Z_4,R)=-e_{\{1\}}^{\emptyset }(R)=-|2\rangle$$
Basis $(2,1)-$vectors (see Eq.(77)), presented as four-component
column vectors with entries indexed by subsets $Z_i$, are
$$
 \left( \begin{array}{c} {e_{\emptyset \vphantom{\{1\}}}^{\emptyset \vphantom{\{2\}}}(Z_1,R)}\\
{e_{\emptyset \vphantom{\{1\}}}^{\emptyset \vphantom{\{2\}}}(Z_2,R)}\\
0\\ 0 \end{array}\right)\
 \left( \begin{array}{c} {e_{\{1\}}^{\{3\}}(Z_1,R)}\\ 0\\ 0\\ {e_{\{1\}}^{\{3\}}(Z_4,R)}
 \end{array}\right)\
 \left( \begin{array}{c} 0\\ e_{\{1\}}^{\{4\}}(Z_2,R)\\ 0\\
 e_{\{1\}}^{\{4\}}(Z_4,R)\end{array}\right)\
\left( \begin{array}{c} {e_{\{2\}}^{\{3\}}(Z_1,R)}\\ 0\\
e_{\{2\}}^{\{3\}}(Z_3,R)\\ 0\end{array}\right)\
$$
$$\left( \begin{array}{c} 0\\ e_{\{2\}}^{\{4\}}(Z_2,R)\\
e_{\{2\}}^{\{4\}}(Z_3,R)\\ 0 \end{array}\right)\
\left(\begin{array}{c} 0\\ 0\\ e_{\{1,2\}}^{\{3,4\}}(Z_3,R)\\
e_{\{1,2\}}^{\{3,4\}}(Z_4,R)\end{array}\right)
$$

Each basis vector (77) corresponds to disassembled determinant,
and, consequently, it is uniquely determined by any of its
non-zero $q-$components. Let us suppose that  $(p+q)-$element
subsets $Z\subset N$ are listed in some fixed ordering (e.g.,
lexical) and let $Z_J^I$ be the subset of {\it the minimal rank}
satisfying the condition $Z\supset (R\backslash J)\cup I$. Then
the relation
$$e_{J'}^{I'}(Z_{J'}^{I'},R)\star
e_{J''}^{I''}(Z_{J''}^{I''},R)=$$ $$=\cases{(-1)^{|(J''\cup
I'')\cap {\Delta}_{(J'\cup I')}|}e_{J'\cup J''}^{I'\cup
I''}(Z_{J'\cup J''}^{I'\cup I''},R) &if $J'\cap J''=\emptyset$ and
$I'\cap I''=\emptyset$\cr 0 &if $J'\cap J''\ne \emptyset \ $ or $\
I'\cap I''\ne \emptyset$\cr}\eqno(78)$$ completely determines star
product of any two basis $(p,q)-$vectors.

For example, to get the product of the second and the fifth
four-com-ponent vectors listed above, it is sufficient to calculate
$$e_{\{1\}}^{\{3\}}(Z_1,R)\star
e_{\{2\}}^{\{4\}}(Z_2,R)=-e_{\{1,2\}}^{\{3,4\}}(Z_3,R)$$ and then
in zero four-component vector insert
$-e_{\{1,2\}}^{\{3,4\}}(Z_i,R)$ in all positions corresponding to
$Z_i\supset \{3,4\}$.

Thus, if some ordering of $(p+q)-$element subsets of $N$ is fixed
and for each admissible $J$ and $I$ subset $Z_J^I$ of minimal
rank, satisfying the condition  $Z\supset (R\backslash J)\cup I$,
is selected then we can consider the following model of the
algebra ${\cal S}_R^{p,q}$: It is a vector space of {\it formal
linear combinations} of $q-$electron vectors $e_J^I(Z_J^I,R)$ with
star product defined by Eq.(78). {\it Note that this model
requires additional structure of linear ordering of all}
$(p+q)-${\it element subsets of} $N$.

 For arbitrary $(p,q)-$vector
$\left \{x_q^Z\right \}_{Z\subset N}$ its $q-$components may be
written in the form
$$x_q^Z=\sum\limits_{{R\backslash (R\cap Z)\subset J\subset
R}\atop{{I\subset Z\backslash (R\cap
Z)}}}x_J^Ie_J^I(Z,R)\eqno(79)$$ where $|J|=|I|$ and {\it the
coefficients} $x_J^I$ {\it do not depend on} $Z$. If, on the other
hand, we have arbitrary family constituted by vectors (79), then
this family is a $(p,q)-$vector because independence of
coefficients in expansion (79) on $Z$ is equivalent to the
conditions (58).

For calculation of $q-$component of star product of two arbitrary
$(p,q)$-vectors the following formula may be used:
$$ {\pi}_{p,q}^Z\left (
\left \{ x_q^{Z'} \right \}_{Z'\subset N}\star \left
\{y_q^{Z'}\right \}_{Z'\subset N} \right) =
\sum\limits_{{R\backslash (R\cap Z)\subset J\subset
R}\atop{{I\subset Z\backslash (R\cap Z)}}}\langle
e_J^I(R)|x_p\star y_p\rangle e_J^I(Z,R).\eqno(80)$$ Here
$$\langle e_J^I(R)|x_p\star y_p\rangle
=\sum\limits_{k_1=0}^{k}\sum\limits_{J_1\subset J\atop{I_1\subset
I }}^{(k_1)}(-1)^{k_1+|(J_1\cup I_1)\cap {\Delta}_{(J\cup
I)}|}x_{J_1}^{I_1}y_{J\backslash J_1}^{I\backslash
I_1},\eqno(81a)$$
$$x_p=s_{q\uparrow p}\left (\left \{ x_q^{Z'} \right \}_{Z'\subset
N}\right ),\ \  y_p=s_{q\uparrow p}\left (\left \{ y_q^{Z'} \right
\}_{Z'\subset N}\right ),\eqno(81b)$$ and $k=|J|=|I|$.

In our previous paper \cite {Panin-3} it was shown that
$(p,q)-$vectors and reduced density operators of order $q$ are
closely related as seen from the equality
$${\rho}_q=\frac{1}{{{n-p}\choose {q}}}\sum\limits_{Z\subset N}^{(p+q)}|x_q^Z\rangle \langle
x_q^Z|\eqno(82)$$ Thus, disassembling mapping  may be interpreted
as a certain {\it pre-contraction operation} applied to
$p-$electron state.  Actual contraction is defined as
$$x_p\to s_{p\downarrow q}(x_p)\to \frac{1}{{{n-p}\choose {q}}}
\sum\limits_{Z\subset N}^{(p+q)}|{\pi}_{p,q}^Zs_{p\downarrow
q}(x_p)\rangle \langle {\pi}_{p,q}^Zs_{p\downarrow
q}(x_p)|\eqno(83)$$ It is pertinent to mention that the
representation of density operator in the form of Eq.(82) is
connected with the commonly used one by a certain non-degenerate
transformation described in detail in paper \cite {Panin-4}.

 Eq.(82) {\it means that any representable by pure $p-$electron
state density operator of order $q$ is a convex hull of pure
normalized $q-$electron states} $v_q^Z\in {\cal F}_{Z,q}$
$${\rho}_q=\sum\limits_{Z\subset N}^{(p+q)}
{\lambda}_Z^2|v_q^Z\rangle \langle v_q^Z|\eqno(84)$$ {\it such
that the family} $\left \{{\lambda}_Zv_q^Z \right \}_{Z\subset N}$
{\it is a $(p,q)-$vector}. This statement, being interesting by
itself, may serve as a base for different parametrizations of
density operators. For example, let us consider ${{n}\choose {q}}$
vectors constituting an orthonormal basis of ${\cal F}_{N,q}$:
$$x_q(i)=\sum\limits_{S\subset N}^{(q)}c_{Si}|S \rangle \eqno(85)$$
and by the symbol $x_q^Z(i)$ let us denote its $Z-$component
$$x_q^Z(i)=\sum\limits_{S\subset Z}^{(q)}c_{Si}|S\rangle
\eqno(86)$$
$(p,q)-$vector
$$\left \{ d_q^{Z} \right \}_{Z\subset N}=
s_{p\downarrow q}\left (C_R|R\rangle
+\sum\limits_{i=1}^{{{n}\choose {q}}}{\mu}_i\sum\limits_{Z'\subset
N}^{(p+q)}*\left (x_q^{Z'}(i)\right )\right )\eqno(87)$$ depending
on $\frac{1}{2}{{n}\choose {q}}\left [{{n}\choose {q}}+1\right ]$
parameters can be used to parametrize representable by pure
$p-$electron states density operators of order $q$. Here
$|R\rangle $ is relevant HF reference state. Note that use of
parametrized density operators associated with parametric
$(p,q)-$vector (87) leads to {\it genuine} $q-$electron
optimization problem but in a certain $p-$electron metric (see
\cite {Panin-3}).

\bigbreak \bigbreak

\bigbreak \bigbreak

{\Large \bf Conclusion} \bigbreak \bigbreak

Any new representation of the space of state vectors
 and especially any new algebraic structure on this space are of
great interest by themselves and deserve thorough study even if
perspectives of their immediate application are vague. In our
opinion with new structure revealed on the Fock space the
situation is different. Its study  clarifies and simplifies the
mathematical backgrounds of CC and related approaches. And even
more, it leads to many relations that would be very difficult to
derive using conventional technique. Eqs.(38), (47), and (53)
supply us with convincing examples of validity of this statement.

The situation with models of the Fock space as  vector spaces of
$(p,q)-$vec-tors is not so unambiguous. These models originate
from the representabilty theory and at first glance are very
complicated. As an immediate application of these models we can
point out possibility to construct parametric {\it representable}
density operators of arbitrary order in a very general way. Of
course, the crucial question arising is the following: what part
of the set of all representable density operators is covered by
such parametrizations? This question, which is probably of
moderate mathematical interest, will has some sense for quantum
chemists only if computer implementation of our approach  will
lead to perspective methods for electronic structure calculations.
The work on implementation of computational methods based on
relations of the type of Eq.(87) is in progress now.

\bigbreak \bigbreak

{ \bf ACKNOWLEDGMENTS} \bigbreak \bigbreak

The author  gratefully acknowledges the Russian Foundation for
Basic Research (Grant 03-03-32335a) for financial support of the
present work.
 \bigbreak

\bigbreak

\end{document}